\documentclass[a4paper,twoside,reqno]{bjp}
\usepackage{graphicx}
\usepackage{cite}
\usepackage{amssymb,amsmath,amscd,amsthm}
\usepackage{times}

\usepackage[bookmarks=false]{hyperref}
\hypersetup{%
    colorlinks=true,        
    linkcolor=blue,          
    citecolor=blue,         
    urlcolor=blue           
    }

\usepackage{geometry}
\usepackage[varg]{txfonts}   
\usepackage{graphicx}
\usepackage{float}
\usepackage{adjustbox}
\usepackage{epstopdf}
\usepackage[version=3]{mhchem}
\usepackage{braket}
\usepackage{latexsym}
\usepackage{float}
\usepackage{subfigure}
\usepackage{young}
\usepackage[dvipsnames]{xcolor}
\usepackage[toc,page]{appendix}
\usepackage{bbm}
\usepackage{tikz}
\usetikzlibrary{positioning,shapes,fit,arrows}
\definecolor{myblue}{RGB}{56,94,141}
 
\allowdisplaybreaks

\begin{document}

\title{Unified picture of nucleon pairs playing leading roles in nuclear collectivity}

\runningheads{D. Bonatsos, A. Martinou, I.E. Assimakis, S.K. Peroulis, S. Sarantopoulou, {\it et al.}}{Unified picture of nucleon collective pairs}

\begin{start}{%
\author{Dennis Bonatsos}{1},
\author{Andriana Martinou}{1},
\author{I.E. Assimakis}{1},
\author{S.K. Peroulis}{1},
\author{S. Sarantopoulou}{1},
\author{N. Minkov}{2}


\address{Institute of Nuclear and Particle Physics, National Centre for Scientific Research Demokritos, Aghia Paraskevi, GR-15310 Attiki, Greece}{1}
\address{Institute of Nuclear Research and Nuclear Energy, Bulgarian Academy of Sciences, 72 Tzarigrad Road, 1784 Sofia, Bulgaria}{2}

\received{Day Month Year (Insert date of submission)}
}

\begin{Abstract}
Proxy-SU(3) symmetry is an approximation scheme extending the Elliott SU(3) algebra of the sd shell to heavier shells. When introduced in 2017, the approximation had been justified by calculations carried out within the Nilsson model, with nucleon pairs differing by $\Delta K [\Delta N \Delta n_z \Delta \Lambda]=0[110]$ in the Nilsson quantum numbers playing a major role in the approximation.   Recently our group managed to map the cartesian basis of the Elliott SU(3) model onto the spherical shell model basis, fully clarifying the approximations used within the proxy-SU(3) scheme and paving the way for using the proxy-SU(3) approximation in shell model calculations for heavy nuclei. As a by-product, the relation of the 0[110] Nilsson pairs used in proxy-SU(3) to the earlier used de Shalit-Goldhaber pairs $|\Delta n \Delta l \Delta j \Delta m_j\rangle = | 0110\rangle$ in spherical shell model notation is clarified, while the Federman-Pittel (FP) pairs known to play a major role at the onset of deformation are identified as $| 0010\rangle$ pairs, and the FP pairs  further increasing the deformation after its onset are found to be  $| 0110\rangle$ pairs, i.e. identical to the de-Shalit-Goldhaber pairs. The connection between the proxy-SU(3) scheme and the spherical shell model has also been worked out in the original framework of the Nilsson model, with identical results.

\end{Abstract}

\begin{KEY}
Elliott SU(3), proxy-SU(3), shell model, Nilsson model, de Shalit-Goldhaber pairs, Federman-Pittel pairs.
\end{KEY}
\end{start}


\section{Bases of main nuclear structure models}

The fundamental underlying model of nuclear structure is the shell model, introduced \cite{Mayer1,Mayer2,Jensen,MJ} in 1949, which is based on a three-dimensional (3D) isotropic harmonic oscillator (HO), with a spin-orbit term added to it. The states are labeled by the number of oscillator quanta $n$, the orbital angular momentum $l$, the total angular momentum $j$, and its $z$-projection $m_j$, with the notation $\vert n l j m_j\rangle$ being used. The shell model was considered adequate for describing near-spherical nuclei, with few valence nucleons outside closed shells. 

In 1952 the collective model of Bohr and Mottelson has been introduced \cite{Bohr,BM}, in order to accommodate  departure from the spherical shape and from axial symmetry,  described by the collective variables $\beta$ and $\gamma$ respectively. 

In 1955, Nilsson  \cite{Nilsson1,NR} introduced a modified version  of the shell model, allowing for axial nuclear deformation to be included, based on a 3D anisotropic HO with cylindrical symmetry \cite{Takahashi,Asherova,RD,ND,PVI,Lenis}. In the Nilsson model the states are labeled by the total number of oscillator quanta $N$, the number $n_z$ of quanta along the $z$-axis, and the projections along the $z$-axis of the orbital angular momentum and the total angular momentum, represented by $\Lambda$ and $K$ respectively. Thus the Nilsson states are denoted by \cite{Nilsson1,NR} $K [N n_z \Lambda]$.

In 1958 Elliott proved that deformation within the nuclear sd shell with U(6) symmetry can be described in terms of its SU(3) subalgebra \cite{Elliott1,Elliott2,Elliott3,Elliott4,Harvey}. In this model the cartesian basis of a 3D isotropic HO is used, $[n_z n_x n_y m_s]$, in which the number of quanta along the $z$, $x$, $y$ directions and the $z$-projection of the spin appear.

In 1972 the group theoretical structure of the Bohr--Mottelson model, having an overall U(5) symmetry possessing an O(5) subalgebra, has been understood \cite{AR}. The seniority quantum number $\tau$, related to the irreducible representations (irreps) of O(5), plays a major role in labeling the states.  

In 1975 the Interacting Boson Model \cite{AI,IA,IVI,FVI} has been introduced, which also has an overall U(6) symmetry built by $s$-bosons of zero angular momentum and $d$-bosons of angular momentum two, possessing three limiting symmetries, U(5) for vibrational nuclei, which is equivalent to the Bohr-Mottelson collective model, O(6) for $\gamma$-unstable nuclei, and SU(3) for deformed nuclei. 

A comprehensive review of using the SU(3) symmetry in nuclear structure has been given recently by Kota \cite{Kota}.  

Beyond the sd nuclear shell the SU(3) symmetry of the 3D isotropic HO is known to be broken by the spin-orbit force, which within each HO shell pushes the orbitals possessing the highest angular momentum $j$ to the shell below. As a consequence, each shell consists by the orbitals left back after this removal, called the normal parity orbitals, plus the orbitals invading from the shell above, having the opposite parity and called the intruder orbitals. 

In 2017 the proxy-SU(3) symmetry has been introduced \cite{proxy1,proxy2,proxy3}, in which the intruder orbitals in each shell (except the one with the highest projection of the total angular momentum) are replaced by the orbitals which have deserted this shell by sinking into the shell below. As a result of this replacement, each shell regains the relevant U(n) symmetry having a SU(3) subalgebra, with only one orbital (which can accommodate two particles) remaining estranged. However, this orbital is the one lying highest in energy within the shell, thus it should be empty for most of the nuclei living in this shell. Therefore it is expected that its influence on the structure of most nuclei living in the shell should be minimal.

The proxy-SU(3) scheme has been initially justified as a good approximation through calculations \cite{proxy1} carried out within the Nilsson model \cite{Nilsson1,NR}. Within the proxy-SU(3) scheme the importance of the highest weight irreducible representations of SU(3) has been demonstrated \cite{EPJAHW} and used \cite{proxy2,proxy3} for the successful prediction of the prolate to oblate shape transition at $N=114$, the dominance of prolate over oblate shapes in the ground states of even-even nuclei, and the prediction of specific islands on the nuclear chart in which shape coexistence can appear \cite{EPJASC}. In the present work we are going to discuss its justification through its connection to the shell model. However, before doing so, it is instructive to discuss the nature of nucleon pairs related to the development of nuclear deformation.

\section{Nucleon pairs favoring deformation} 

As early as 1953 it has been observed by deShalit and Goldhaber \cite{deShalit} in their studies of $\beta$ transition probabilities that within the proton--neutron pairs of orbitals (1p3/2, 1d5/2), (1d5/2, 1f7/2), (1f7/2,   1g9/2), (1g9/2, 1h11/2), (1h11/2, 1i13/2) the nucleons of one kind (protons, for example) have a stabilizing effect on pairs of nucleons of the other kind (neutrons in the example), thus favoring the development of nuclear deformation. In the standard shell model notation $\vert n l j m_j\rangle$, the orbitals within each pair of parentheses differ by   $\vert \Delta n \Delta l \Delta j \Delta m_j\rangle = \vert 0 1 1 0\rangle$. 

A major step forward in our understanding of effective interactions and coupling schemes in nuclei has been taken in 1962 by Talmi \cite{Talmi62} through the introduction of seniority \cite{Talmi62,Talmi71,Talmi73,Talmi93}, representing the number of nucleon pairs coupled to non-zero angular momentum, which explained the linear dependence of neutron separation energies on the mass number within series of isotopes.

\begin{table}[htbp]
\caption{The transformation matrix $R\cdot C$ for $N=2$. The shell model orbitals appear in the first line, while the Elliott orbitals appear in the first column. These orbitals are used in the harmonic oscillator shell 8-20 ($sd$ shell), or in the proxy-SU(3) shell 14-26 after the replacement of the intruder orbitals with their de Shalit--Goldhaber partners. Adapted from Ref. \cite{EPJASM}.}\small\smallskip
\tabcolsep=3.6pt

\begin{tabular}{@{}ccccccc@{}}
\hline
\hline
&&&&&&\\[-8pt]

$\begin{array}{ccccccc}

\ket{n,l,j,m_j} & \ket{2s^{1/2}_{-1/2}} & \ket{2s^{1/2}_{1/2}} & \ket{1d^{3/2}_{-3/2}} & \ket{1d^{3/2}_{-1/2}} &  \ket{1d^{3/2}_{1/2}} & \ket{1d^{3/2}_{3/2}} \\

\ket{n_z,n_x,n_y,m_s}& & & & & & \\ 

\noalign{\smallskip}\hline\noalign{\smallskip}

\ket{0,0,2,-\frac{1}{2}} & -\frac{1}{\sqrt{3}} & 0 & 0 & -\frac{1}{\sqrt{15}} & 0 & -\frac{1}{\sqrt{5}} \\
 
\ket{0,0,2,\frac{1}{2}}& 0 & -\frac{1}{\sqrt{3}} & \frac{1}{\sqrt{5}} & 0 & \frac{1}{\sqrt{15}} & 0 \\

\ket{0,1,1,-\frac{1}{2}}& 0 & 0 & 0 & 0 & 0 & -i \sqrt{\frac{2}{5}} \\

\ket{0,1,1,\frac{1}{2}}& 0 & 0 & -i \sqrt{\frac{2}{5}} & 0 & 0 & 0\\

\ket{0,2,0,-\frac{1}{2}}& -\frac{1}{\sqrt{3}} & 0 & 0 & -\frac{1}{\sqrt{15}} & 0 & \frac{1}{\sqrt{5}}\\

\ket{0,2,0,\frac{1}{2}}& 0 & -\frac{1}{\sqrt{3}} & -\frac{1}{\sqrt{5}} & 0 & \frac{1}{\sqrt{15}} & 0 \\

\ket{1,0,1-\frac{1}{2}}& 0 & 0 & \frac{i}{\sqrt{10}} & 0 & i \sqrt{\frac{3}{10}} & 0 \\

\ket{1,0,1,\frac{1}{2}}& 0 & 0 & 0 & -i \sqrt{\frac{3}{10}} & 0 & -\frac{i}{\sqrt{10}}\\

\ket{1,1,0,-\frac{1}{2}}& 0 & 0 & \frac{1}{\sqrt{10}} & 0 & -\sqrt{\frac{3}{10}} & 0 \\

\ket{1,1,0,\frac{1}{2}}& 0 & 0 & 0 & -\sqrt{\frac{3}{10}} & 0 & \frac{1}{\sqrt{10}} \\

\ket{2,0,0,-\frac{1}{2}}& -\frac{1}{\sqrt{3}} & 0 & 0 & \frac{2}{\sqrt{15}} & 0 & 0 \\

\ket{2,0,0,\frac{1}{2}}& 0 & -\frac{1}{\sqrt{3}} & 0 & 0 & -\frac{2}{\sqrt{15}} & 0 \\

\end{array} $

\end{tabular} 

\bigskip\bigskip

\begin{tabular}{@{}ccccccc@{}}

$\begin{array}{ccccccc}

\ket{n,l,j,m_j} & \ket{1d^{5/2}_{-5/2}}& \ket{1d^{5/2}_{-3/2}} & \ket{1d^{5/2}_{-1/2}} & \ket{1d^{5/2}_{1/2}} & \ket{1d^{5/2}_{3/2}} & \ket{1d^{5/2}_{5/2}}\\

\ket{n_z,n_x,n_y,m_s}& & & & & & \\

\noalign{\smallskip}\hline\noalign{\smallskip}

\ket{0,0,2,-\frac{1}{2}}&  -\frac{1}{2} & 0 & -\frac{1}{\sqrt{10}} & 0 & -\frac{1}{2 \sqrt{5}}& 0 \\

\ket{0,0,2,\frac{1}{2}}&  0 & -\frac{1}{2 \sqrt{5}} & 0 & -\frac{1}{\sqrt{10}} & 0 & -\frac{1}{2}\\

\ket{0,1,1,-\frac{1}{2}}& \frac{i}{\sqrt{2}} & 0 & 0 & 0 & -\frac{i}{\sqrt{10}} & 0 \\

\ket{0,1,1,\frac{1}{2}}&  0 & \frac{i}{\sqrt{10}} & 0 & 0 & 0 & -\frac{i}{\sqrt{2}} \\

\ket{0,2,0,-\frac{1}{2}}&  \frac{1}{2} & 0 & -\frac{1}{\sqrt{10}} & 0 & \frac{1}{2 \sqrt{5}}& 0 \\

\ket{0,2,0,\frac{1}{2}}&  0 & \frac{1}{2 \sqrt{5}} & 0 & -\frac{1}{\sqrt{10}} & 0 & \frac{1}{2}\\

\ket{1,0,1-\frac{1}{2}}&  0 & i \sqrt{\frac{2}{5}} & 0 & \frac{i}{\sqrt{5}} & 0 & 0 \\

\ket{1,0,1,\frac{1}{2}}&  0 & 0 & \frac{i}{\sqrt{5}} & 0 & i \sqrt{\frac{2}{5}} & 0 \\

\ket{1,1,0,-\frac{1}{2}}&  0 & \sqrt{\frac{2}{5}} & 0 & -\frac{1}{\sqrt{5}} & 0 & 0 \\

\ket{1,1,0,\frac{1}{2}}&  0 & 0 & \frac{1}{\sqrt{5}} & 0 & -\sqrt{\frac{2}{5}} & 0 \\

\ket{2,0,0,-\frac{1}{2}}&  0 & 0 & \sqrt{\frac{2}{5}} & 0 & 0 & 0 \\

\ket{2,0,0,\frac{1}{2}}&  0 & 0 & 0 & \sqrt{\frac{2}{5}} & 0 & 0 \\

\end{array}$

\end{tabular}
\label{N2}
\end{table}

In 1977 Federman and Pittel \cite{FP1,FP2,FP3} realized that when adding valence protons and valence neutrons to a nucleus, the proton--neutron pairs (1d5/2, 1d3/2), (1g9/2, 1g7/2), (1h11/2, 1h9/2), and (1i13/2, 1i11/2) are responsible for the onset of deformation, while deformation is then established by the proton--neutron pairs (1d5/2, 1f7/2), (1g9/2, 1h11/2), (1h11/2, 1i13/2), and (1i13/2, 1j15/2).
These sets correspond to $\vert \Delta n \Delta l \Delta j \Delta m_j\rangle = \vert 0 0 1 0\rangle$ and $\vert 0 1 1 0\rangle$ respectively in the shell model notation, the latter set coinciding with the de 
Shalit--Goldhaber pairs. 

The decisive role played by proton-neutron pairs has been demonstrated in 1985 through the introduction of the $N_p N_n$ scheme \cite{CastenPRL,Casten} and the $P$-factor, $P= N_p N_n / (N_p+N_n)$ \cite{Haustein,Castenbook}, by showing the systematic dependence of several observables on the competition between the quadrupole deformation, ``measured'' by the quadrupole-quadrupople interaction through $N_pN_n$, and the pairing interaction, ``measured'' through $N_p+N_n$, where $N_p$ ($N_n$) is the number of valence protons (neutrons). 

In 1995 the quasi-SU(3) symmetry \cite{Zuker1,Zuker2} has been introduced, based on the proton--neutron pairs (1g9/2, 2d5/2), (1h11/2, 2f7/2), (1i13/2, 2g9/2), expressed as $\vert \Delta n \Delta l \Delta j \Delta m_j\rangle = \vert 1 2 2 0\rangle$ in the shell model notation, which lead to enhanced quadrupole collectivity \cite{Kaneko}. 

Following detailed studies of double differences of binding energies \cite{Cakirli94,Cakirli96,Brenner,Stoitsov,Cakirli102}, in 2010 it has been realized \cite{Burcu} that proton-neutron pairs differing in the Nilsson notation \cite{Nilsson1,NR} $K [N n_z \Lambda]$ by $\Delta K[ \Delta N \Delta n_z \Delta \Lambda]=0[110]$, play a major role in the development of nuclear deformation, due to their large spatial overlaps \cite{Sofia}. These pairs correspond to the replacements made within the proxy-SU(3) scheme \cite{proxy1,proxy2,proxy3}. No relation to the pairs mentioned in the previous paragraphs had been realized at that time.


\begin{table}[htbp]
\caption{The shell model orbitals of the original spin-orbit like shells and of the proxy-SU(3) shells. The magic number $14$ is proposed as a sub-shell closure in Ref. \cite{Sorlin}. The symmetry of each proxy-SU(3) shell is U($\Omega$) with $\Omega={({N}+1)({N}+2)\over 2}$ \cite{Kota}. The orbitals being replaced are denoted by bold letters. Adapted from Ref. \cite{EPJASM}.}\small\smallskip

\tabcolsep=3.6pt
\begin{tabular}{@{}ccccc@{}}
\hline
\hline
&&&&\\[-8pt]

$U(\Omega)$ & original orbitals & proxy orbitals  &  SO magic numbers & HO magic numbers \\ 
 \noalign{\smallskip}\hline\noalign{\smallskip}
 
U(3)      & $1p^{1/2}_{\pm 1/2}$ & $1p^{1/2}_{\pm 1/2}$ & 6-14 & 2-8 \smallskip\\
          &  $\bf 1d^{5/2}_{\pm 1/2,\pm 3/2}$&  $\bf 1p^{3/2}_{\pm 1/2,\pm 3/2}$ & &  \smallskip\\
          & $\bf 1d^{5/2}_{\pm 5/2}$ & - & & \bigskip\\
 
U(6)      & $2s^{1/2}_{\pm 1/2}$ & $2s^{1/2}_{\pm 1/2}$ & 14-28 & 8-20  \smallskip\\
          & $1d^{3/2}_{\pm 1/2,\pm 3/2}$ & $1d^{3/2}_{\pm 1/2,\pm 3/2}$  & & \smallskip\\
          & $\bf 1f^{7/2}_{\pm 1/2,\pm 3/2,\pm 5/2}$ & $\bf 1d^{5/2}_{\pm 1/2,\pm 3/2,\pm 5/2}$ & & \smallskip\\
          & $\bf 1f^{7/2}_{\pm 7/2}$ & - & & \bigskip\\

U(10)     & $2p^{1/2}_{\pm 1/2}$ & $2p^{1/2}_{\pm 1/2}$  & 28-50 & 20-40 \smallskip\\
          & $2p^{3/2}_{\pm 1/2,\pm 3/2}$ &  $2p^{3/2}_{\pm 1/2,\pm 3/2}$  & & \smallskip\\
          & $1f^{5/2}_{\pm 5/2, \pm3/2,\pm 1/2}$ & $1f^{5/2}_{\pm 5/2, \pm3/2,\pm 1/2}$  & & \smallskip\\
          & $\bf 1g^{9/2}_{\pm 1/2,..., \pm 7/2}$ & $\bf 1f^{7/2}_{\pm 1/2,..., \pm 7/2}$  & & \smallskip\\
          & $\bf 1g^{9/2}_{\pm 9/2}$ & - & & \bigskip\\

U(15)     & $3s^{1/2}_{\pm 1/2}$ &  $3s^{1/2}_{\pm 1/2}$ & 50-82 & 40-70 \smallskip\\
          & $2d^{3/2}_{\pm 1/2,\pm 3/2}$ & $2d^{3/2}_{\pm 1/2,\pm 3/2}$  & & \smallskip\\
          & $2d^{5/2}_{\pm 1/2,...,\pm 5/2}$ &  $2d^{5/2}_{\pm 1/2,...,\pm 5/2}$ & &  \smallskip\\
          & $1g^{7/2}_{\pm 1/2,...,\pm 7/2}$ & $1g^{7/2}_{\pm 1/2,...,\pm 7/2}$ & & \smallskip\\
          & $\bf 1h^{11/2}_{\pm 1/2,...,\pm 9/2}$ & $\bf 1g^{9/2}_{\pm 1/2,...,\pm 9/2}$  & & \smallskip\\
          & $\bf 1h^{11/2}_{\pm 11/2}$ & -  & & \bigskip\\
 
U(21)     & $3p^{1/2}_{\pm 1/2}$ & $3p^{1/2}_{\pm 1/2}$ & 82-126 & 70-112  \smallskip\\
          & $3p^{3/2}_{\pm 1/2,\pm 3/2}$ & $3p^{3/2}_{\pm 1/2,\pm 3/2}$ & & \smallskip\\
          & $2f^{5/2}_{\pm 1/2,...,\pm 5/2}$ & $2f^{5/2}_{\pm 1/2,...,\pm 5/2}$  & & \smallskip\\
          & $2f^{7/2}_{\pm 1/2,...,\pm 7/2}$ & $2f^{7/2}_{\pm 1/2,...,\pm 7/2}$  & & \smallskip\\
          & $1h^{9/2}_{\pm 1/2,...,\pm 9/2}$ & $1h^{9/2}_{\pm 1/2,...,\pm 9/2}$ & & \smallskip\\
          & $\bf 1i^{13/2}_{\pm 1/2,...,\pm 11/2}$ & $\bf 1h^{11/2}_{\pm 1/2,...,\pm 11/2}$ & &  \smallskip\\
          & $\bf 1i^{13/2}_{\pm 13/2}$ & -  \bigskip \\

\hline
\hline
\end{tabular}
\label{shells}
\end{table}

\begin{table}
\caption{Expansions of Nilsson orbitals $K[N n_z \Lambda]$ in the shell model basis $|N l j m_j \rangle$ for three different values of the deformation $\epsilon$. The Nilsson orbitals shown possess the highest total angular momentum $j$ in their shell. The existence of a leading shell model eigenvector is evident at all deformations.  Adapted from Ref. \cite{EPJPSM}. } \small\smallskip

\tabcolsep=3.6pt
\begin{tabular}{@{}rcrrrrrr@{}}
\hline
\hline
&&&\\[-8pt]

${3\over 2}[541]$ & $\epsilon$ & $|N l j m_j \rangle$ & 
$\left| 5 1 {3\over 2} {3\over 2} \right\rangle$ & 
$\left| 5 3 {5\over 2} {3\over 2} \right\rangle$ & 
$\left| 5 3 {7\over 2} {3\over 2} \right\rangle$ & 
$\left| 5 5 {9\over 2} {3\over 2} \right\rangle$ & 
$\left| 5 5 {11\over 2} {3\over 2} \right\rangle$ \\

\noalign{\smallskip}\hline\noalign{\smallskip}
 & 0.05 & & 0.0025 & $-0.0015$ & 0.0641 & $-0.0122$ & 0.9979 \\
 & 0.22 & & 0.0371 & $-0.0286$ & 0.2565 & $-0.0640$ & 0.9633 \\
 & 0.30 & & 0.0601 & $-0.0506$ & 0.3287 & $-0.0922$ & 0.9366 \\

\noalign{\smallskip}\hline
\end{tabular}

\begin{tabular}{ r c  r r r r r r r }
\hline\noalign{\smallskip}
${3\over 2}[651]$ & $\epsilon$ & $|N l j m_j \rangle$ & 
$\left| 6 2 {3\over 2} {3\over 2} \right\rangle$ & 
$\left| 6 2 {5\over 2} {3\over 2} \right\rangle$ & 
$\left| 6 4 {7\over 2} {3\over 2} \right\rangle$ & 
$\left| 6 4 {9\over 2} {3\over 2} \right\rangle$ & 
$\left| 6 6 {11\over 2} {3\over 2} \right\rangle$ &
$\left| 6 6 {13\over 2} {3\over 2} \right\rangle$ \\

\noalign{\smallskip}\hline\noalign{\smallskip}

 & 0.05 & & $-0.0002$ & 0.0046 & $-0.0013$ & 0.0821 & $-0.0086$ & 0.9966  \\
 & 0.22 & & $-0.0100$ & 0.0711 & $-0.0278$ & 0.3240 & $-0.0469$ & 0.9418  \\
 & 0.30 & & $-0.0207$ & 0.1149 & $-0.0509$ & 0.4091 & $-0.0687$ & 0.9010  \\

\hline
\hline
\end{tabular}
\label{Hadi1}
\end{table}

\begin{table}[htbp]
\caption{Expansions of Nilsson orbitals $K[N n_z \Lambda]$ in the shell model basis $|N l j m_j \rangle$ for three different values of the deformation $\epsilon$. The Nilsson orbitals shown do not possess the highest total angular momentum $j$ in their shell. The existence of a leading shell model eigenvector is evident at small deformation, but this is not the case any more at higher deformations, at which several shell model eigenvectors make considerable contributions. Adapted from Ref. \cite{EPJPSM}.} \small\smallskip

\tabcolsep=3.6pt
\begin{tabular}{@{}rcrrrrrr@{}}
\hline
\hline
&&&\\[-8pt]

${1\over 2}[431]$ & $\epsilon$ & $|N l j m_j \rangle$ & 
$\left| 4 0 {1\over 2} {1\over 2} \right\rangle$ & 
$\left| 4 2 {3\over 2} {1\over 2} \right\rangle$ & 
$\left| 4 2 {5\over 2} {1\over 2} \right\rangle$ & 
$\left| 4 4 {7\over 2} {1\over 2} \right\rangle$ & 
$\left| 4 4 {9\over 2} {1\over 2} \right\rangle$ \\

\noalign{\smallskip}\hline\noalign{\smallskip}
 & 0.05 & & $-0.0213$ & 0.1254 & $-0.0702$ & 0.9893 & 0.0127 \\
 & 0.22 & & $-0.2248$ & 0.4393 & $-0.2791$ & 0.8057 & 0.1717 \\
 & 0.30 & & $-0.2630$ & 0.5003 & $-0.2458$ & 0.7447 & 0.2559 \\

\noalign{\smallskip}\hline
\end{tabular}

\begin{tabular}{ r c r r r r r r r }
\hline\noalign{\smallskip}
${1\over 2}[541]$ & $\epsilon$ & $|N l j m_j \rangle$ & 
$\left| 5 1 {1\over 2} {1\over 2} \right\rangle$ & 
$\left| 5 1 {3\over 2} {1\over 2} \right\rangle$ & 
$\left| 5 3 {5\over 2} {1\over 2} \right\rangle$ & 
$\left| 5 3 {7\over 2} {1\over 2} \right\rangle$ & 
$\left| 5 5 {9\over 2} {1\over 2} \right\rangle$ &
$\left| 5 5 {11\over 2} {1\over 2} \right\rangle$ \\

\noalign{\smallskip}\hline\noalign{\smallskip}

 & 0.05 & & $-0.0200$ & 0.1770 & $-0.0295$ & 0.9780 & $-0.0446$ & $-0.0944$  \\
 & 0.22 & & $-0.2492$ & 0.4619 & $-0.3768$ & 0.5550 & $-0.4161$ & $-0.3185$  \\
 & 0.30 & & $-0.3121$ & 0.4331 & $-0.4829$ & 0.3430 & $-0.4789$ & $-0.3671$  \\
 
\hline
\hline
\end{tabular}
\label{Hadi2}
\end{table} 

\section{Connecting the cartesian Elliott basis to the spherical shell model basis} 

The cartesian basis of Elliott \cite{Elliott1,Elliott2,Elliott3,Elliott4,Harvey} can be transformed to the spherical basis $[n l m_l m_s]$ in $l$-$s$ coupling through a unitary transformation \cite{Davies,Chasman,Chacon} 
\begin{equation}
[n_z n_x n_y m_s] = R [n l m_l m_s], 
\end{equation}
the details of which can be found in Ref. \cite{EPJASM}. Using Clebsch-Gordan coefficients  the spherical basis can be rewritten in $j$-$j$ coupling as 
\begin{equation}
[n l m_l m_s] =  C [n l j m_j],  
\end{equation}
in which the total angular momentum $j$ and its $z$-projection appear. Combining these two transformations one obtains 
\begin{equation}
[n_z n_x n_y m_s] = R  C [n l j m_j],  
\end{equation}
i.e., the connection between the cartesian Elliott basis and the shell model basis in $j$-$j$ coupling. An example of this transformation is shown in Table \ref{N2}. Details of the calculations and tables for other shells can be found in Ref. \cite{EPJASM}. 

Using the above transformation one sees that the Nilsson 0[110] replacements made within the proxy-SU(3) scheme are ``translated'' into $| 0 1 1 0\rangle$ replacements within the spherical shell model basis. The resulting correspondence between  original shell model orbitals and proxy-SU(3) orbitals is summarized in Table \ref{shells}. This correspondence proves that the 0[110] Nilsson pairs identified in Ref. \cite{Burcu} and used within the proxy-SU(3) scheme \cite{proxy1,proxy2,proxy3} are identical to the de Shalit--Goldhaber pairs \cite{deShalit} and the Federman--Pittel pairs \cite{FP1,FP2,FP3} within the spherical shell model basis, in which they are expressed as $| 0 1 1 0\rangle$ pairs. 

The correspondence between Nilsson pairs and shell model pairs has been corroborated by calculations \cite{EPJPSM} within the Nilsson model, in which the first justification of the proxy-SU(3) scheme has been found \cite{proxy1}. As one can see in Table \ref{Hadi1}, the correspondence used in proxy-SU(3) works well for the Nilsson orbitals which possess the highest total angular momentum $j$ within their shell, which are exactly the orbitals which are replaced within the proxy-SU(3) scheme, while it breaks down for Nilsson orbitals which do not possess the highest total angular momentum $j$ within their shell, as seen in Table \ref{Hadi2}. In further corroboration of this result, a unitary transformation connecting the orbitals being replaced within the proxy-SU(3) scheme has been found \cite{EPJASM} within the shell model basis.

This correspondence also paves the way for taking advantage of the proxy-SU(3) symmetry in shell model calculations for heavy nuclei, in a way similar to that of the symmetry-adapted no-core shell model approach \cite{Launey1,Launey2} used in light nuclei.

\section{The importance of the basis choice}

As an example of the drastic modification of convergence in the presence of an underlying symmetry one can recall the Algebraic Collective Model of D.J. Rowe et al. \cite{Rowe735,Rowe753,JPA38,Caprio,Welsh,RW}. The solution of the Bohr Hamiltonian  becomes very much time consuming if one goes to deformed nuclei. This is because the original Bohr Hamiltonian has the U(5)$\supset$SO(5)$\supset$SO(3) structure, thus convergence is expected to be rapid for vibrational nuclei, but time consuming for deformed nuclei. But taking advantage of the underlying SU(1,1) symmetry of the modified harmonic oscillator \cite{Bahri} (like the Davidson potential \cite{Davidson}) one can greatly reduce the size of the calculation for deformed nuclei, by selecting a basis based on the SU(1,1)$\times$SO(5) symmetry. In this  basis only one vector will make the most important contribution in the case of a deformed state \cite{Rowe735,Rowe753,JPA38,Caprio,Welsh}, while the other vectors will have minimal contributions. 

From the above it becomes clear that the choice of an appropriate basis is of crucial importance for keeping numerical calculations tractable. In the case of the shell model, the traditional approaches work well in light nuclei, as well as in heavy nuclei very close to the magic numbers (i.e., very close to closed shells), while calculations for heavy nuclei away from the magic numbers become intractable.  It is hoped that the reinstatement of the SU(3) symmetry in heavy nuclei away from closed shells would lead to rapid convergence. One way in which the necessary reinstatement of the 3D harmonic oscillator SU(3) symmetry can be achieved is through the introduction of the proxy-SU(3) symmetry.

\section*{Acknolwledgements} 

{Financial support by the Greek State Scholarships Foundation (IKY) and the European Union within the MIS 5033021 action, as well as by the Bulgarian National Science Fund (BNSF) under Contract No.KP-06-N48/1  is gratefully acknowledged.}

\end{document}